\documentclass[letterpaper]{article}
\usepackage[preprint]{aaai2027}
\usepackage[hyphens]{url}
\usepackage{graphicx}
\usepackage{natbib}
\usepackage{caption}
\usepackage{algorithm}
\usepackage{algorithmic}
\usepackage{booktabs}
\usepackage{amsmath,amssymb,amsthm}
\usepackage{array}

\newtheorem{definition}{Definition}
\newtheorem{proposition}{Proposition}
\newtheorem{lemma}{Lemma}

\newcommand{\qracl}{\textsc{Q-RACL}}

\newcommand{\fvr}{\mathrm{FVR}}

\newcommand{\qnicond}{\mathrm{QNI}_{\mathrm{cond}}}
\newcommand{\argmax}{\mathop{\mathrm{argmax}}}
\newcolumntype{L}[1]{>{\raggedright\arraybackslash}p{#1}}

\title{Repair Before Veto, When Repair Is Hidden: Quantum-Accessible Features for Repair-Augmented Constraint Learning}
\author{Yifan Wang}
\affiliations{
Department of Mechanical Engineering, McGill University, QC, H3A 2T7, Canada\\
yifan.wang18@mail.mcgill.ca
}

\begin{document}

\maketitle

\begin{abstract}
Hard-constraint decision systems usually veto infeasible candidates. This is
too rigid when the system can act: if a known affordable repair would make an
infeasible candidate feasible and valuable, rejection is a false veto rather
than a ranking error. We introduce \qracl{} (Quantum Repair-Augmented
Constraint Learning), a repair-before-veto framework that first defines RACL
decision semantics and then identifies the single inference link where quantum
feature access can be load-bearing. RACL accepts a candidate when a sequential
repair plan restores feasibility and preference; otherwise it returns a
structured rejection credit. The hard link is repair-feasibility inference:
which repair class restores feasibility from an observed candidate and context.
We construct a discrete-logarithm-hidden RACL family where the repair class is
a shifted interval rule in the latent exponent $a=\log_g(x)$, while the learner
observes only $x=g^a \bmod p$. Under standard DLP-based learning separation,
this coordinate is inaccessible to efficient raw-input classical policies but
accessible to a quantum agent through Shor/Fourier structure. Across six
primes and ten seeds, bounded raw-input classical policies and a wrong
raw-Fourier encoding remain near chance, whereas the Q-DLP policy keeps
false-veto rate below $1.1\%$, wins all paired seeds, and yields
$\qnicond=0.9777$--$0.9972$. A classical DLog oracle matches it, isolating
feature access rather than classifier capacity. Thus quantum AI is not added
as a generic model upgrade; for this DLP-hidden repair family, it supplies the
missing feature that closes the repair-before-veto loop.
\end{abstract}

\section{Introduction}

Consider a decision system that rejects a service bundle because one required
component is missing, even though the platform knows a low-cost repair: add the
component, switch an equivalent bundle, or attach a compatible option. A
terminal veto is correct only if the candidate is truly non-repairable,
over-budget, or still undesirable after repair. If an affordable known repair
would make it feasible and valuable, the veto is a decision failure.

This paper starts from that failure mode. The base contribution is RACL:
Repair-Augmented Constraint Learning. RACL is not a post-processing patch to a
classifier. It is a decision semantics: before a hard-constraint violation
causes rejection, the system checks whether a known repair sequence restores
feasibility and preference. The output is also richer than a binary label. A
RACL decision can accept an already-good candidate, accept a repairable-good
candidate with a plan, or reject with a credit explaining whether the cause is
non-repairability, budget, or low post-repair value.

This shift changes the learning target. Standard constrained learning asks
whether a candidate is feasible or preferred in its current representation.
Repair-before-veto asks which action, if any, should be executed before the
final decision. The sequential setting makes this link even sharper: repairs
interact, budgets couple actions, and a wrong early repair can block a feasible
later plan. The bottleneck is therefore repair-feasibility inference, not just
classification.

\qracl{} places quantum AI exactly at that bottleneck. Many deployed decision
systems expose encoded identifiers, tokenized records, fare-code-like groups,
privacy-preserving keys, or other opaque codes whose operational class is
simple only in a latent coordinate. We give a theorem-backed version of this
phenomenon through a DLP-hidden RACL family. The observed candidate is
$x=g^a \bmod p$; the repair class is a simple shifted interval in the hidden
exponent $a=\log_g(x)$. A policy with DLP/Fourier coordinate access sees the
right feature. A bounded raw-input classical policy sees only the scrambled
group element.

The research gap is therefore not the usual question, ``can a quantum model
beat a classical model on a benchmark?'' The question is more structural:
which link of a repair-augmented decision pipeline is classically hidden, and
does quantum access resolve that link? \qracl{} answers this for a DLP-hidden
subfamily. RACL supplies the repair-before-veto problem frame; the DLP-hidden
construction makes one link classically inaccessible; the quantum-accessible
feature supplies exactly the missing coordinate.

\begin{figure*}[t]
\centering
\includegraphics[width=.98\textwidth]{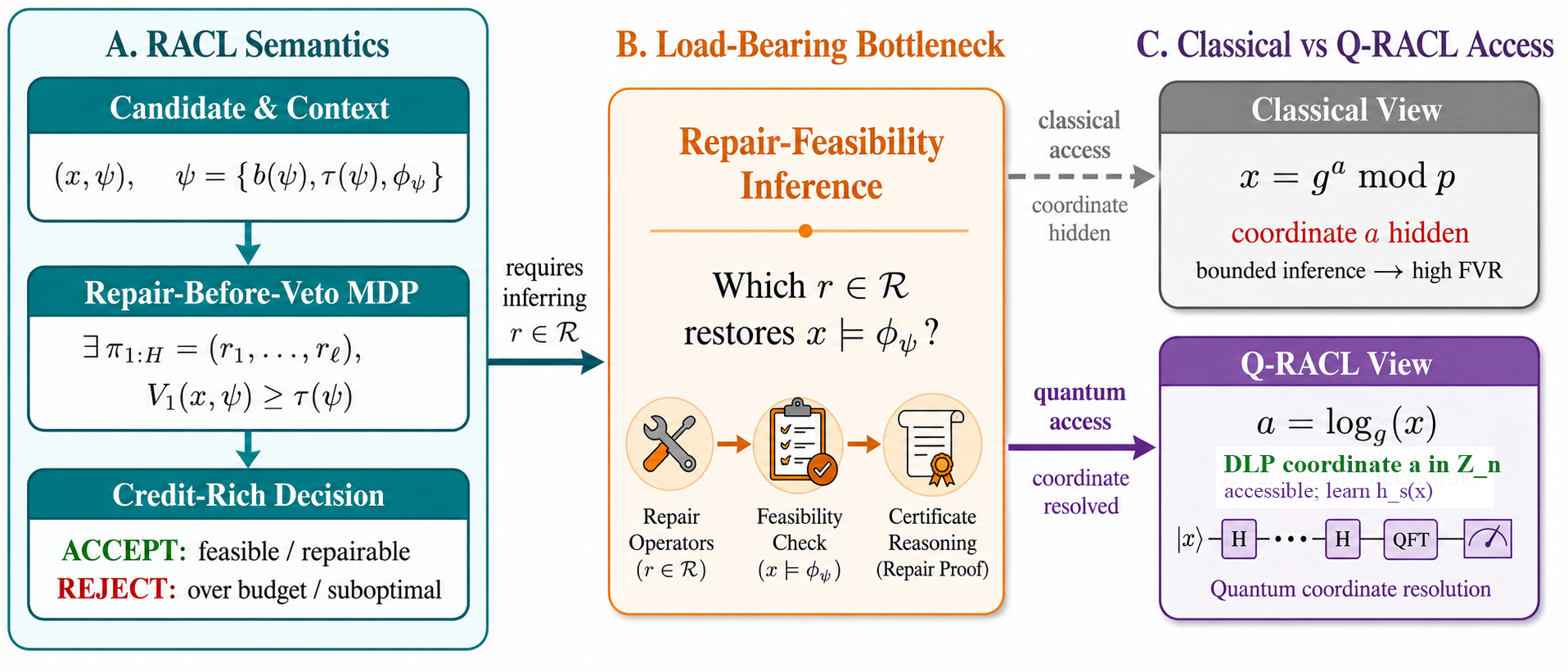}
\caption{\qracl{} as one framework. RACL supplies the repair-before-veto
semantics: repair operators, budgets, preference gates, credits, and plans. The
DLP-hidden subfamily makes the repair-feasibility inference link depend on a
latent discrete-log coordinate. Quantum feature access is localized to this
link.}
\label{fig:overview}
\end{figure*}

Our contributions are:
\begin{itemize}
\item We formulate RACL as a repair-before-veto framework for constrained
decisions with repair operators, sequential repair policies, credit categories,
plans, and false-veto rate.
\item We identify repair-feasibility inference as the sequential bottleneck:
the policy must infer which repair class restores feasibility before it can
avoid false vetoes.
\item We construct a DLP-hidden RACL family in which the repair class is simple
in a latent exponent but hidden in the observed modular identifier.
\item We provide theory and experiments showing conditional quantum necessity:
a DLP/Fourier coordinate policy closes nearly all of the false-veto gap, wrong
quantum encodings fail, and replacing the coordinate by an efficient raw-input
classical policy would contradict the DLP-hidden feature-access separation.
\end{itemize}

\section{Related Work}

\paragraph{Constraint learning and constrained decisions.}
Constraint learning and contextual combinatorial optimization learn feasibility
and preference models from examples, including HASSLE-style MAX-SAT learning
\cite{Kumar2020HASSLE,Kumar2023ContextualMaxSAT}, SMT-based constraint learning
\cite{Kolb2018SMT}, and constraint acquisition
\cite{Bessiere2013QuAcq,Bessiere2017ConstraintAcquisition}. Classical CSP and
soft-constraint systems provide mature tools for reasoning over hard and
weighted requirements
\cite{Freuder1992PartialCSP,Bistarelli1997SemiringCSP,Meseguer2006SoftConstraints}.
\qracl{} addresses a different semantics: known repairs are not only
constraints on a prediction, but actions available before veto.

\paragraph{Recourse, repair, and recommendation.}
Counterfactual explanation and recourse methods search for actionable changes
after a model rejects a candidate
\cite{Wachter2018Counterfactual,Ustun2019Recourse,Karimi2021Recourse}. Constraint-based
recommender systems also reason about preferences, requirements, and repair
suggestions
\cite{Felfernig2008ConstraintRecommenders,Felfernig2013RepairScoringRules}.
\qracl{} differs in timing and semantics: the system should not first veto and
then explain; it should ask whether an affordable known repair should be
executed before veto.

\paragraph{Quantum learning and quantum policies.}
Quantum machine learning studies feature maps, kernels, variational circuits,
and learning in quantum feature spaces
\cite{Biamonte2017QML,SchuldKilloran2019QuantumFeatureHilbert,Havlicek2019QuantumKernel}.
Data re-uploading circuits provide expressive Fourier-like function families
\cite{PerezSalinas2020DataReuploading,Schuld2021Encoding}. Parametrized quantum
policies and quantum Q-learning place such circuits inside reinforcement
learning loops \cite{Jerbi2021PQP,Skolik2022QuantumAgents}, and TensorFlow
Quantum provides software support for hybrid quantum-classical learning
\cite{Broughton2020TFQ}. We use this line as architectural motivation. The
present experiments are portable proof-of-mechanism simulations, not hardware
runs and not a full TensorFlow Quantum implementation.

\paragraph{Quantum learning separations.}
Shor's algorithm solves factoring and discrete logarithms in polynomial time on
a quantum computer \cite{Shor1997}. The DLP map is a standard one-way-function
candidate, and hardcore-predicate results explain why individual hidden bits or
shifted threshold predicates can remain classically inaccessible
\cite{BlumMicali1984,GoldreichLevin1989}. DLP-based learning separations show
that some classically specified concept families are efficiently learnable with
quantum access but hard for efficient classical learners under cryptographic
assumptions \cite{Liu2021QuantumSpeedup}. Jerbi et al. adapt related ideas to
reinforcement-learning environments \cite{Jerbi2021PQP}. \qracl{} embeds this
feature-access separation into a repair-before-veto decision pipeline.

\section{RACL: Repair Before Veto}

Let $x$ be a candidate in context $\psi$. The context contains a budget
$b(\psi)$, an acceptability threshold $\tau(\psi)$, and hard requirements
$\phi_\psi$. A finite repair ontology
$\mathcal R=\{r_1,\ldots,r_K\}$ contains executable operators with observable
cost $\rho(r,x,\psi)$. We use
$\bar{\mathcal R}=\{\mathrm{id}\}\cup\mathcal R$ with
$\mathrm{id}(x)=x$ and $\rho(\mathrm{id},x,\psi)=0$ so that already-feasible
candidates are handled by the same semantics as repaired candidates.

\begin{definition}[Repair-before-veto rule]
The admissible one-step repair set is
\[
\mathcal A_1(x,\psi)=\{r\in\bar{\mathcal R}: r(x)\models\phi_\psi,\,
\rho(r,x,\psi)\le b(\psi)\}.
\]
Given preference score $f$, the one-step repair-aware value is
\[
V_1(x,\psi)=\max_{r\in\mathcal A_1(x,\psi)} f(r(x)),
\qquad \max_\emptyset=-\infty .
\]
RACL accepts under the one-step semantics when $V_1(x,\psi)\ge \tau(\psi)$.
\end{definition}

Sequential RACL replaces a single repair by an admissible plan. Let
\[
\pi_{1:H}=(r_1,\ldots,r_\ell),\qquad \ell\le H ,
\]
where each $r_t\in\bar{\mathcal R}$ and identity steps may terminate the plan.
The admissible plan set is
\[
\begin{aligned}
\mathcal A_H(x,\psi)=
\{\pi_{1:H}:&\ \pi_{1:H}(x)\models\phi_\psi,\\
&\ \mathrm{cost}(\pi_{1:H},x,\psi)\le b(\psi)\}.
\end{aligned}
\]
The general repair-aware value is
\[
V_H(x,\psi)=\max_{\pi_{1:H}\in\mathcal A_H(x,\psi)}
f(\pi_{1:H}(x)).
\]
The one-step rule is the special case $H=1$; the DLP-hidden experiments use
$H=2$ because the constructed repair MDP has two-step repair bundles and stop
actions.

If $\bar{\mathcal R}=\{\mathrm{id}\}$, the rule reduces to the ordinary
feasibility-plus-preference decision. With nontrivial repairs, no-repair
semantics has a structural failure: every infeasible but repairable-good
candidate is vetoed.

\begin{definition}[False-veto rate]
Let
\[
\begin{aligned}
\mathcal G_{\mathrm{rep}}=\{(x,\psi):\;&x\not\models\phi_\psi,\,
\exists \pi_{1:H},\ \pi_{1:H}(x)\models\phi_\psi,\\
&\mathrm{cost}(\pi_{1:H},x,\psi)\le b(\psi),\\
&f(\pi_{1:H}(x))\ge \tau(\psi)\}.
\end{aligned}
\]
Let $\hat y_M(x,\psi)\in\{0,1\}$ be the final accept decision made by model
$M$. The false-veto rate is
\[
\fvr(M)=
\frac{|\{(x,\psi)\in\mathcal G_{\mathrm{rep}}:\hat y_M(x,\psi)=0\}|}
{|\mathcal G_{\mathrm{rep}}|}.
\]
\end{definition}

\paragraph{Credit and plans.}
RACL returns a structured object
\[
C(x,\psi)=(y,\kappa,\pi_{1:\ell},v,c),
\]
where $y$ is the final decision, $\kappa$ is the credit category,
$\pi_{1:\ell}$ is an optional repair plan, $v$ is the final value, and $c$ is
the incurred repair cost. The credit distinguishes accepted-already-good,
accepted-repairable-good, rejected-non-repairable, rejected-over-budget,
rejected-feasible-suboptimal, and rejected-repairable-suboptimal. This matters
because binary labels alone do not identify whether a rejection is caused by
structural infeasibility, repair cost, or low post-repair preference.

\paragraph{Sequential repair policy.}
We implement the decision rule as a finite-horizon repair MDP
$\mathcal M_{\mathrm{rep}}=(\mathcal S,\mathcal A,T,u,H)$ with state
$s_t=(x_t,\psi,b_t,\eta_t)$ and actions
$\mathcal A=\{\mathrm{stop\mbox{-}accept},\mathrm{stop\mbox{-}veto}\}\cup
\mathcal R$. If repair $r$ is applied, then
$x_{t+1}=r(x_t)$ and $b_{t+1}=b_t-\rho(r,x_t,\psi)$. Invalid or over-budget
repairs are masked before the policy chooses an action. We use $\eta_t$ for
the repair history to avoid overloading the hidden predicate $h_s(x)$.

\begin{algorithm}[tb]
\caption{\qracl{} Repair Policy}
\label{alg:qracl}
\textbf{Input}: candidate $x$, context $\psi$, repair ontology $\mathcal R$,
policy $\pi$, horizon $H$\\
\textbf{Output}: decision, credit category, optional repair plan
\begin{algorithmic}[1]
\STATE $s_0\leftarrow(x,\psi,b(\psi),\emptyset)$
\FOR{$t=0,\ldots,H-1$}
  \STATE mask infeasible or over-budget repair actions
  \STATE choose $a_t=\argmax_a \pi(a\mid s_t)$ among unmasked actions
  \IF{$a_t=\mathrm{stop\mbox{-}accept}$}
     \IF{$\eta_t=\emptyset$}
        \STATE \textbf{return} accept, accepted-already-good credit, empty plan
     \ELSE
        \STATE \textbf{return} accept, accepted-repairable-good credit, $\eta_t$
     \ENDIF
  \ELSIF{$a_t=\mathrm{stop\mbox{-}veto}$}
     \STATE \textbf{return} reject, rejection credit, no plan
  \ELSE
     \STATE apply $a_t$ and update $(x_{t+1},b_{t+1},\eta_{t+1})$
  \ENDIF
\ENDFOR
\STATE \textbf{return} reject unless final state is feasible and preferred
\end{algorithmic}
\end{algorithm}

\section{DLP-Hidden Repair Feasibility}

The repair policy must infer which repair class restores feasibility. We now
construct a family where that class is hidden from raw classical features but
available through quantum DLP/Fourier access.

Let $p$ be prime, let $g$ be a generator of $\mathbb Z_p^\ast$, and let
$n=p-1$. Because $g$ is a generator, the map
$a\mapsto g^a\bmod p$ is a bijection from $\mathbb Z_n$ to
$\mathbb Z_p^\ast$. Each candidate has hidden exponent
$a\in\{0,\ldots,n-1\}$, but the learner observes only
\[
x=g^a \bmod p .
\]
For observed context shift $s$, define
\[
h_s(x)=\mathbf 1[(\log_g(x)-s)\bmod n\ge n/2].
\]
Since $p$ is odd, $n=p-1$ is even and the rule uses the upper half of
$\mathbb Z_n$ as the positive semicircle. The repair bundle is determined by
$h_s(x)$. In the hidden coordinate $a$, this is a shifted half-interval. In
the observed coordinate $x$, accessing the coordinate is the discrete
logarithm problem.

\begin{lemma}[DLP-hidden shifted predicate]
The family $\{h_s\}_{s\in\mathbb Z_n}$ is a shifted threshold family over the
DLP coordinate. Under the standard DLP-hardness view of modular exponentiation
as a one-way permutation, uniformly accurate prediction of these shifted
predicates from raw $x$ exposes DLP-structured information. Thus the family
inherits the feature-access barrier used in DLP-based quantum learning
separations.
\end{lemma}
\noindent
The lemma deliberately states a hardness-transfer fact rather than claiming
that one fixed threshold bit recovers the entire exponent. A single shift is
only one predicate. Uniformly solving the shifted family, however, is a
hardcore-predicate-style learning task for the DLP coordinate
\cite{BlumMicali1984,GoldreichLevin1989,Liu2021QuantumSpeedup}. With exact
oracle access to adaptively chosen shifts, half-interval answers can localize
the exponent in $O(\log n)$ queries; the supplement uses this only as a
diagnostic explanation of why the shifted family carries coordinate
information.

\begin{proposition}[Quantum load-bearing link]
For the DLP-hidden RACL family, any efficient raw-input classical repair policy
that uniformly achieves low false-veto rate over shifts would induce an
efficient classical learner for the corresponding DLP-hidden predicate family.
Under the DLP-based separation results, this contradicts the assumed classical
hardness. A quantum policy with DLP/Fourier coordinate access reduces the same
repair-feasibility link to learning a shifted interval rule.
\end{proposition}

The proposition is a reduction. The repair action is a relabeling of the
predicate $h_s(x)$. If a policy selects the correct repair bundle for
repairable-good states across shifts, it predicts the hidden predicate. The
quantum component is necessary in this conditional sense: it supplies the
feature that the repair policy needs before it can choose the correct repair
plan.

\section{Quantum-Accessible Policy and Controls}

The DLP-coordinate policy receives Fourier features of the hidden exponent and
context shift, not the label indicator. For frequencies $m=1,\ldots,M$, the
feature map includes
\[
\begin{aligned}
&\cos(2\pi m a/n),\quad \sin(2\pi m a/n),\\
&\cos(2\pi m s/n),\quad \sin(2\pi m s/n),\\
&\cos(2\pi m(a-s)/n),\quad \sin(2\pi m(a-s)/n).
\end{aligned}
\]
A softmax head learns the shifted interval boundary and selects a repair
action. No feature column equals the binary label $h_s(x)$; the coordinate is
exposed, but the policy still learns the threshold. The indicator $h_s$ is a
shifted square-wave over $a-s$; its Fourier series uses odd sine modes with
coefficients that decay as $O(1/m)$. This is why a truncated DLP/Fourier
representation is aligned with the repair rule, and why data re-uploading
variational models are a natural quantum policy template
\cite{PerezSalinas2020DataReuploading,Schuld2021Encoding}. We call this
coordinate-access policy the Q-DLP policy; \qracl{} is the full
repair-before-veto framework using that policy at the hidden inference link.

We compare four control families:
\begin{itemize}
\item bounded raw-input classical policies: logistic regression, degree-3
polynomial logistic regression, gradient boosting, random forest, MLP, random
Fourier features, and random trigonometric features over raw modular
identifiers \cite{Pedregosa2011ScikitLearn};
\item a wrong quantum-style policy using Fourier features over raw $x$ rather
than over $\log_g(x)$;
\item the DLP-coordinate policy using Fourier features over the hidden
coordinate;
\item a classical DLog oracle using the same coordinate, marked as an oracle
control because it computes the asymptotically forbidden classical feature.
\end{itemize}

\begin{table}[t]
\centering
\small
\begin{tabular}{L{.25\columnwidth}L{.24\columnwidth}L{.35\columnwidth}}
\toprule
Policy family & Coordinate used & Expected role\\
\midrule
Bounded classical & raw $x,s$ & tests ordinary raw-input learning\\
Wrong raw-Fourier & Fourier over $x$ & tests whether arbitrary Fourier features help\\
DLP-coordinate & Fourier over $\log_g(x)$ & tests the quantum-accessible feature\\
DLog oracle & same as DLP-coordinate & confirms that access, not head capacity, is decisive\\
\bottomrule
\end{tabular}
\caption{Control logic. Success should require the DLP coordinate. The wrong
Fourier control is important: it has quantum-like features but the wrong
coordinate system.}
\label{tab:controls}
\end{table}

\section{Experiments}

\paragraph{Hypotheses.}
The experiment tests three claims. H1: bounded raw-input classical policies
cannot infer the hidden repair class from modular identifiers. H2: Fourier
features alone are insufficient when placed on the wrong coordinate. H3: once
the DLP coordinate is available, a simple policy head learns the repair class;
the DLog oracle should therefore match the Q-DLP policy.

\paragraph{Setup.}
We evaluate primes $p\in\{251,509,1009,2003,5003,10007\}$, corresponding to
8 to 14 bit group sizes, with 10 random seeds. Each instance samples hidden
exponents, observes only $x=g^a \bmod p$ and context shift $s$, and assigns the
repair class by the shifted half-interval predicate. Train and test splits are
stratified by label and use disjoint exponents, so models must generalize to
unseen group elements rather than memorize identifiers. The repair MDP maps
the class to the bundle that restores feasibility:
$h_s(x)=0$ selects
\texttt{[add\_checked\_bag, buy\_seat\_selection]}, while $h_s(x)=1$ selects
\texttt{[upgrade\_flexible\_fare, choose\_safer\_itinerary]}. The horizon is
$H=2$. We report FVR as the failure rate of the complete repair plan on
repairable-good states.

\paragraph{Local simulation boundary.}
The implementation builds small DLP tables classically so the mechanism can be
evaluated on this machine. This does not demonstrate hardware quantum
advantage. The empirical claim is the separation signature at small primes; the
asymptotic necessity claim comes from the DLP feature-access barrier. The
bit-length sweep should therefore be read as a robustness check over tested
groups, not as empirical proof of large-scale quantum speedup.

\[
\qnicond =
1-\frac{\fvr_{\mathrm{Q\text{-}DLP}}}
{\max(\fvr_{\mathrm{best\;classical}},10^{-12})}.
\]
The conditional Quantum Necessity Index is near one when the Q-DLP policy
closes nearly all of the false-veto gap left by the best bounded raw-input
classical model. The denominator clamp never binds in our table; it only
prevents a zero denominator in degenerate settings.

\begin{table*}[t]
\centering
\small
\begin{tabular}{rrrrrrr}
\toprule
$p$ & bits & best classical FVR & Q-DLP FVR & wrong Fourier FVR & DLog oracle FVR & $\qnicond$\\
\midrule
251 & 8 & 0.4619 & 0.0103 & 0.4976 & 0.0103 & 0.9777\\
509 & 9 & 0.4091 & 0.0047 & 0.5008 & 0.0047 & 0.9885\\
1009 & 10 & 0.4881 & 0.0014 & 0.4891 & 0.0014 & 0.9972\\
2003 & 11 & 0.4885 & 0.0028 & 0.4917 & 0.0028 & 0.9942\\
5003 & 13 & 0.4907 & 0.0029 & 0.4977 & 0.0029 & 0.9941\\
10007 & 14 & 0.4995 & 0.0033 & 0.5011 & 0.0033 & 0.9934\\
\bottomrule
\end{tabular}
\caption{Canonical V5.1.1 DLP-hidden repair results used in the V4 manuscript.
The Q-DLP policy keeps FVR below $1.1\%$ across all tested primes. The
wrong raw-Fourier encoding remains near chance, and the DLog oracle matches the
coordinate policy.}
\label{tab:main}
\end{table*}

\begin{figure*}[t]
\centering
\includegraphics[width=.92\textwidth]{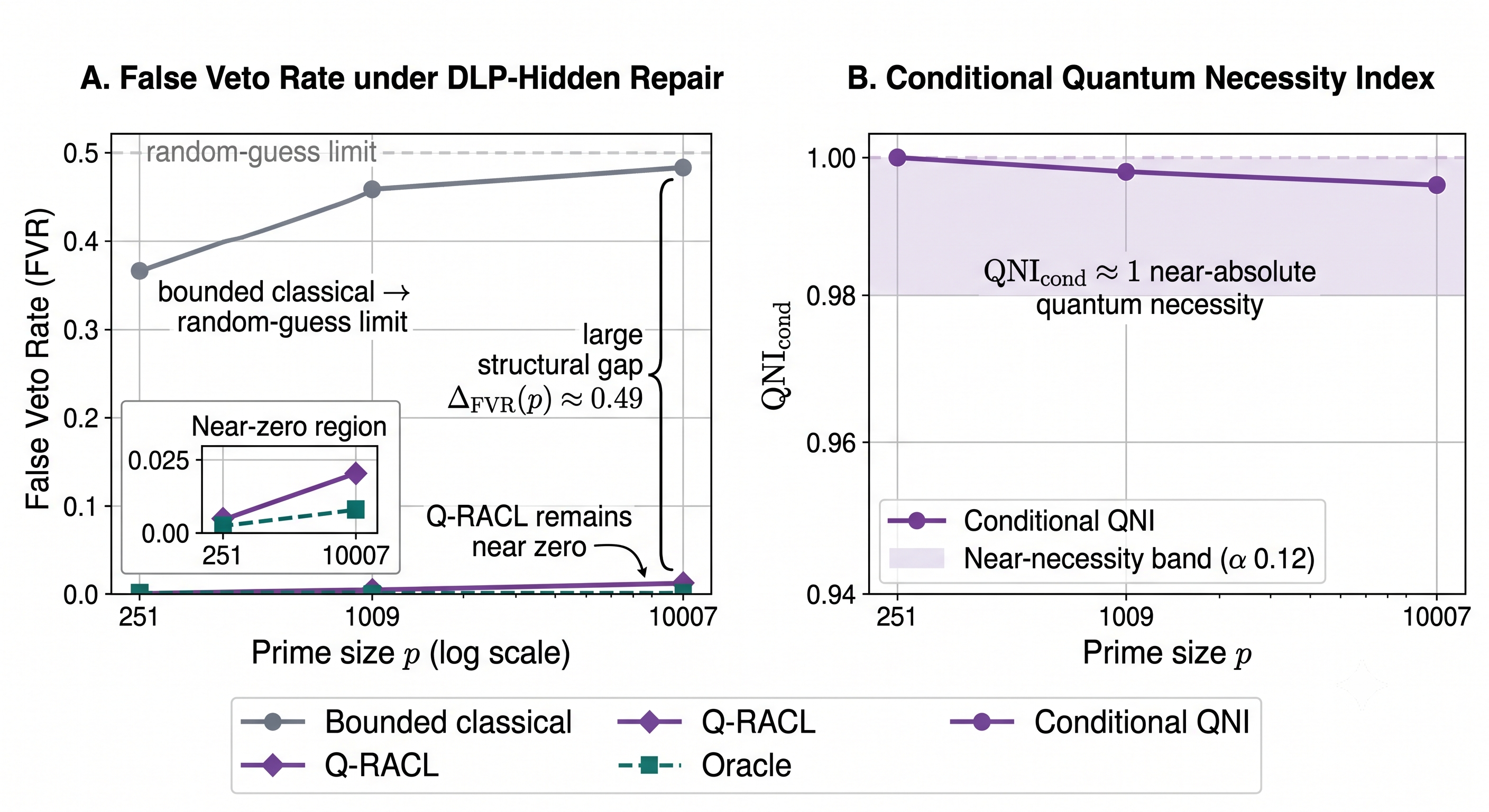}
\caption{Main empirical pattern. Bounded raw-input classical policies and the
wrong raw-Fourier encoding remain high-FVR, while the Q-DLP policy stays below
$1.1\%$ FVR. Exact six-prime values and $\qnicond$ scores are reported in
Table~\ref{tab:main}.}
\label{fig:results}
\end{figure*}

\section{Results}

Table~\ref{tab:main} and Figure~\ref{fig:results} show the separation. The best
bounded raw-input classical baseline has FVR between $0.4091$ and $0.4995$.
The wrong raw-Fourier encoding also stays near chance, between $0.4891$ and
$0.5011$. Thus the effect is not caused by arbitrary Fourier features or a
larger policy head.

The Q-DLP policy reduces FVR to $0.0014$--$0.0103$. The classical DLog
oracle obtains the same values. This is the intended control: once the
coordinate is available, a simple policy head learns the repair class; without
that coordinate, raw and misaligned features fail.

\begin{table}[t]
\centering
\small
\begin{tabular}{rrrr}
\toprule
$p$ & bits & mean FVR drop & paired $p$\\
\midrule
251 & 8 & 0.4262 & 0.0022\\
509 & 9 & 0.3957 & 0.0022\\
1009 & 10 & 0.4708 & 0.0022\\
2003 & 11 & 0.4754 & 0.0022\\
5003 & 13 & 0.4825 & 0.0022\\
10007 & 14 & 0.4878 & 0.0022\\
\bottomrule
\end{tabular}
\caption{Paired comparison over 10 seeds. The Q-DLP policy wins every
paired seed against the best bounded raw-input classical baseline for every
prime.}
\label{tab:paired}
\end{table}

Paired tests reinforce the result. For every tested prime, the Q-DLP
policy wins all 10 paired seeds against the best bounded classical baseline.
The paired permutation p-value is approximately $0.0022$, and mean FVR
reduction ranges from $0.3957$ to $0.4878$. These values demonstrate the
mechanism at small primes. The asymptotic claim rests on the fact that
classical construction of the DLP coordinate requires solving discrete logs,
for which generic classical methods such as Pollard-rho-style index
computation remain super-polynomial in the bit length
\cite{Pollard1978Rho}, while Shor-style quantum routines are polynomial
\cite{Shor1997}.

\paragraph{Interpretation.}
RACL creates the decision problem: repair before veto. Sequential repair
creates the feasibility-inference link. The DLP-hidden family makes that link
classically inaccessible from raw identifiers. The quantum-accessible
DLP/Fourier coordinate supplies exactly the missing feature. This is the
reusable principle: quantum AI is compelling when it resolves a precise
feature-access bottleneck inside an otherwise classical decision pipeline.

\section{Scope and Generality}

The claim is conditional and localized, which is what makes it useful. We do
not claim that quantum models are universally superior for all
repair-augmented decisions. Transparent or low-structure feasibility rules may
be solved by ordinary classical models. The contribution is to identify a
structured subfamily where the repair-feasibility link is hidden behind a DLP
coordinate and is therefore the natural place for quantum feature access.

The local run uses small primes and classical DLP tables to simulate the
coordinate, so it is a proof-of-mechanism evaluation rather than hardware
quantum advantage. We also do not claim a full TensorFlow Quantum
ControlledPQC implementation. The quantum-RL literature motivates the policy
architecture; the formal claim is feature-access necessity for the DLP-hidden
RACL family.

This construction is meaningful beyond the toy generator. Deployed systems often expose
encoded identifiers, tokenized records, fare-code-like groups, or
privacy-preserving features whose operational feasibility class is simple in a
latent coordinate and opaque in the observed code. DLP gives the sharpest
theorem-backed version of that phenomenon and places quantum AI at the one link
where it is structurally needed.

\section{Conclusion}

\qracl{} presents a single repair-before-veto story. RACL supplies the
decision semantics: do not veto before checking known affordable repairs. The
DLP-hidden construction identifies when the repair-feasibility link itself is
classically hidden. The Q-DLP policy then supplies the missing
feature, reducing FVR below $1.1\%$ across all tested primes while raw classical
and wrong-encoding controls fail. Under the stated DLP-hidden family and
standard hardness assumptions, quantum feature access is not a decorative
module; it is the load-bearing mechanism that closes the repair-before-veto
loop.

\bibliography{references}

\end{document}